\documentclass[sigconf,nonacm]{acmart}
\usepackage{eso-pic}
\renewcommand\footnotetextcopyrightpermission[1]{}
\ifdefined\InternalOnlyHeader
  \AddToShipoutPictureFG{%
    \AtPageUpperLeft{%
      \raisebox{-0.34in}[0pt][0pt]{%
        \makebox[\paperwidth]{\footnotesize\bfseries INTERNAL ONLY - CONFIDENTIAL}%
      }%
    }%
  }%
\fi
\AtBeginDocument{%
  }

\begin{document}

\newcommand{\yyh}[1]{\textcolor{blue}{#1}}

\newcommand{\zyl}[1]{\textcolor{red}{\textbf{#1}}}

\title{Potential Applications of HBF in LLM Serving Systems}

\author{Yihan Yin}
\email{yyhsess2021@stu.pku.edu.cn}
\affiliation{%
  \institution{School of Integrated Circuits\\Peking University\\Alibaba DAMO Academy}
  \country{China}
}

\author{Yilun Zhao}
\email{zyilun8@gmail.com}
\affiliation{%
  \institution{Alibaba DAMO Academy}
  \country{China}
}

\author{Zhixin Yun}
\email{xinzsky6@gmail.com}
\affiliation{%
  \institution{Alibaba DAMO Academy}
  \country{China}
}

\author{Guanying Wu}
\email{wuguanying.wgy@alibaba-inc.com}
\affiliation{%
  \institution{Alibaba Cloud Computing}
  \country{China}
}

\author{Feng Zhu}
\email{f.zhu@alibaba-inc.com}
\affiliation{%
  \institution{Alibaba Cloud Computing}
  \country{China}
}

\author{Kai Tao}
\email{kaitao.tk@alibaba-inc.com}
\affiliation{%
  \institution{Alibaba Cloud Computing}
  \country{China}
}

\author{Shu Li}
\email{s.li@alibaba-inc.com}
\affiliation{%
  \institution{Alibaba Cloud Computing}
  \country{China}
}

\author{Fei Huang}
\email{fangwu.hf@alibaba-inc.com}
\affiliation{%
  \institution{Alibaba DAMO Academy\\Hupan Lab}
  \country{China}
}

\author{Zhe Zhang}
\email{zz225296@alibaba-inc.com}
\affiliation{%
  \institution{Alibaba DAMO Academy\\Hupan Lab}
  \country{China}
}

\author{Shuangchen Li}
\email{shuangchen.li@alibaba-inc.com}
\affiliation{%
  \institution{Alibaba DAMO Academy\\Hupan Lab}
  \country{China}
}

\author{Hongzhong Zheng}
\email{hongzhong.zheng@alibaba-inc.com}
\affiliation{%
  \institution{Alibaba DAMO Academy\\Hupan Lab}
  \country{China}
}

\begin{abstract}
LLM serving is increasingly constrained by memory capacity as model weights,
KV caches, and the number of served model variants continue to grow. This
report examines High-Bandwidth Flash (HBF) as a capacity-oriented extension to
HBM-based serving systems. We first discuss how HBF can be integrated into the
GPU memory hierarchy without undermining the bandwidth expected by the compute
die. We then model the system-level value of added capacity as expanded
residency for read-mostly model-state objects. Under this view, HBF can improve
MoE serving by enabling more expert replicas and can improve multi-model
serving by reducing model loading and supporting hot-model replication. Our
simulation results show that these benefits depend on preserving the
HBM-resident execution path while using HBF to expand the resident set of model
weights.

\end{abstract}

\maketitle

\section{Introduction}
\label{sec:introduction}

Large language models (LLMs) are increasingly used in online applications such as chatbots, coding assistants, and emerging agentic workflows, making cloud LLM serving an important datacenter workload.
Serving these applications demands token generation under stringent latency requirements while accessing massive volumes of data.
The data volume continues to grow as individual models become larger, context lengths increase, and providers deploy broader sets of model variants, increasing the GPU memory capacity required for serving \cite{kwon2023vllm,sheng2023flexgen,duanMuxServeFlexibleSpatialtemporal2024,yuPrismUnleashingGPU2025a}.
At the same time, interactive and agentic workloads make per-token latency increasingly important, requiring the memory system to supply data fast enough for low-latency generation \cite{agrawal2024sarathi,zhongDistServeDisaggregatingPrefill2024a}.
Together, these capacity and bandwidth demands make memory a primary design constraint in LLM serving \cite{kwon2023vllm,sheng2023flexgen,maChallengesResearchDirections2026}.

Modern GPUs rely on high bandwidth memory (HBM) to provide the bandwidth and capacity required by LLM serving.
HBM stacks multiple DRAM dies using through-silicon vias (TSVs) and connects the stack to the compute die through a silicon interposer in a 2.5D package.
This organization combines substantial capacity from vertical DRAM stacking with high bandwidth from the wide interposer-based interface.

\begin{table}[t]
\centering
\caption{Representative HBM stack configurations.}
\label{tab:hbm-scaling}
\vspace{-0.5em}
\begin{tabular}{lcc}
\toprule
HBM stack & Capacity per stack & Bandwidth per stack \\
\midrule
HBM3 12Hi     & 24~GB  & 819~GB/s \\
HBM3E 12Hi    & 36~GB  & \(>1.2\)~TB/s \\
HBM4 12Hi     & 36~GB  & \(>2.8\)~TB/s \\
\bottomrule
\end{tabular}
\vspace{-0.7em}
\end{table}

Across generations, HBM has increased both bandwidth and capacity, as summarized in Table~\ref{tab:hbm-scaling} \cite{micronHBM3E12High36GB2024,micronHBM4ProductPage2026}.
However, these two dimensions do not scale with the same difficulty; while bandwidth has maintained a relatively smooth scaling pace through wider interfaces and higher per-pin data rates, capacity scaling has become increasingly constrained at both the stack and package levels.
At the stack level, increasing capacity typically requires denser DRAM dies or more stacked dies, which increases TSV, bonding, stacking, and thermal-management complexity and can make yield control more difficult.
At the package level, adding more HBM stacks consumes interposer area, dense routing resources, and package power delivery, limiting the number of stacks that can be integrated around the compute die.
Recent products illustrate this asymmetry. Micron's 12-high HBM4 stack keeps the same 36~GB capacity as its HBM3E counterpart, while increasing per-stack bandwidth to more than 2.8~TB/s, more than twice that of HBM3E \cite{micronHBM3E12High36GB2024,micronHBM4ProductPage2026}.
As a result, HBM capacity is becoming the harder dimension to scale for future LLM serving platforms.

To move beyond this HBM capacity-scaling bottleneck, SanDisk has proposed High-Bandwidth Flash (HBF), a NAND-flash-based memory technology for AI inference~\cite{sandiskHBFFactSheet2025}.
HBF follows a stacked-memory organization similar to HBM and uses NAND flash cells to target much higher capacity density while still providing high read bandwidth.
However, turning this capacity advantage into LLM serving benefits raises two questions.
At the architecture level, HBF cannot simply replace HBM because flash introduces longer access latency and coarser access granularity than DRAM-based HBM.
The architecture question is therefore how to organize HBM and HBF so that HBF adds capacity without reducing the effective bandwidth needed for low-latency token generation.
At the system level, existing LLM serving systems are designed under the capacity limits of HBM-only GPUs.
With HBF substantially expanding available memory capacity, the question is how system design should be adapted to better exploit the additional capacity.

\begin{figure}
  \includegraphics[width=\columnwidth]{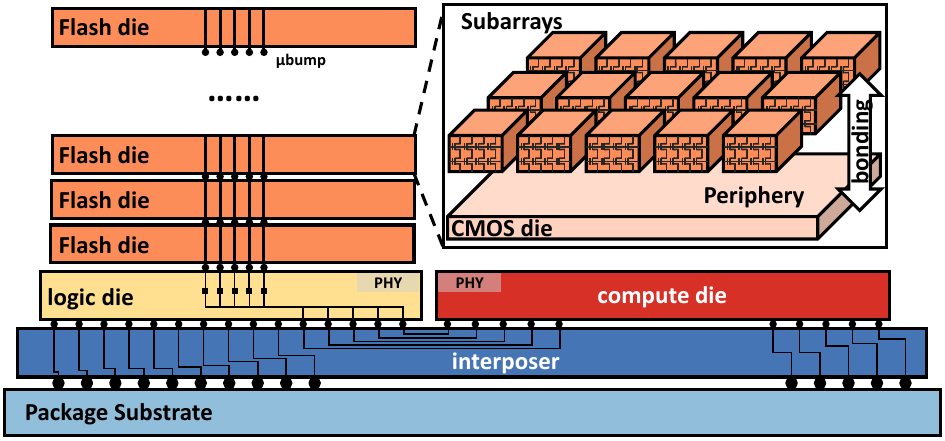}
  \vspace{-1.15em}
  \caption{Conceptual organization of HBF as stacked NAND-flash dies with logic and vertical interconnects packaged near the compute die. Conceptual diagram, not drawn to actual layer proportions.}
  \label{fig:hbf_concept}
  \vspace{-0.5em}
\end{figure}

\section{Motivation}
\label{sec:motivation}

\subsection{Data-Intensive LLM Serving}
\label{sec:motivation-llm-serving}

An LLM request is processed in two phases. The \emph{prefill} phase processes
the input tokens and initializes the key-value (KV) cache, while the subsequent
\emph{decoding} phase generates output tokens autoregressively, with each new
token conditioned on the previously generated tokens. This autoregressive
decoding loop is memory-bound because every output token requires repeated access to
model weights and to the growing KV cache~\cite{zhongDistServeDisaggregatingPrefill2024a}.

Both sources of data are increasing. Longer context windows enlarge the KV
cache associated with each active request, while larger models increase the
weight footprint that must be served. MoE models further amplify the capacity
pressure because the total parameter set can grow much faster than the number
of parameters activated per token. For example, DeepSeek-V3 has 671B total
parameters and activates 37B parameters per token~\cite{deepseek-aiDeepSeekV3TechnicalReport2025};
DeepSeek-V4-Pro scales this footprint to 1.6T total parameters while activating
49B parameters per token~\cite{deepseekaiDeepSeekV4Preview2026}.

\begin{figure}[t]
  \centering
  \includegraphics[width=\columnwidth]{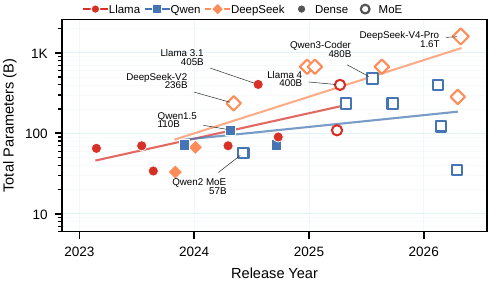}
  \vspace{-1.15em}
  \caption{Representative server-side LLM variants from Llama, Qwen, and DeepSeek.}
  \label{fig:model-scale-variety}
  \vspace{-0.5em}
\end{figure}

The service-level footprint also grows because providers deploy model families.
A family may include general-purpose, code,
vision, reasoning, low-latency, and MoE variants, each targeting a different
capability, latency, context-length, or cost point. Figure~\ref{fig:model-scale-variety}
illustrates this trend for representative server-side variants from the Llama,
Qwen, and DeepSeek families. Over time, these families introduce both larger
models and more specialized dense/MoE variants. Thus, the memory demand of an
LLM service grows along two dimensions, the size of each served model and the
number of variants that must coexist in the serving system.

At the same time, latency requirements remain stringent and are becoming even
tighter for interactive workloads. MLPerf Inference v5.0 includes a low-latency
Llama 2 70B Interactive workload with a 40~ms time-per-output-token (TPOT)
constraint, corresponding to 25 output tokens per second per user~\cite{mlcommonsMLPerfInferenceV50Results2025,nvidiaBlackwellMLPerfInferenceV502025}.
Together, increasing data volume and tight TPOT targets require memory systems
to scale in both bandwidth and capacity.

\subsection{Capacity Constraints in LLM Serving and System-Level Workarounds}
\label{sec:motivation-serving-systems}

Modern GPUs utilize HBM to provide the high bandwidth required for LLM serving.
However, HBM capacity is limited by stack-level and package-level physical constraints.
These constraints necessitate system-level compromises when serving large MoE models or multiple model variants.

For large MoE models, a deployment may need to spread the expert set across many GPUs to fit within aggregate HBM capacity.
Serving systems therefore use expert parallelism to shard experts across GPUs.
DeepSeek-V3, for example, uses 32-way expert parallelism for prefill and 320-way expert parallelism for decoding in its production deployment~\cite{deepseek-aiDeepSeekV3TechnicalReport2025}.
Thus, expert parallelism alleviates HBM capacity pressure by aggregating memory across devices, while introducing dispatch/combine communication whenever tokens are routed to remote experts.

\begin{figure}
    \centering
    \includegraphics[width=\linewidth]{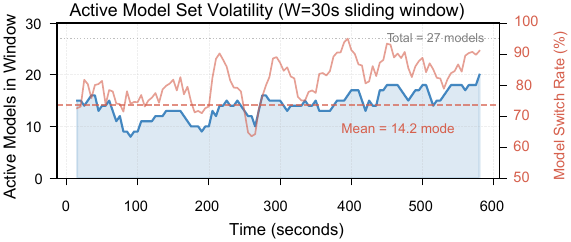}
    \vspace{-1.15em}
    \caption{Temporal variation of active models.}
    \label{fig:volatility}
    \vspace{-0.5em}
\end{figure}

In multi-model serving, capacity pressure comes from keeping all models available under limited HBM capacity.
Providers serve many models with skewed and time-varying popularity. A small set of hot models receives most requests, while requests to many long-tail models are sparse or bursty, as shown in Figure~\ref{fig:volatility}.
Existing systems address this pressure with capacity-aware resource-management strategies. 
They colocate models according to popularity, optimize checkpoint loading or migration for cold models, and pool GPU memory across models to improve utilization under limited HBM capacity~\cite{duanMuxServeFlexibleSpatialtemporal2024,fuServerlessLLMLowLatencyServerless2024a,xiangAegaeonEffectiveGPU2025a,yuPrismUnleashingGPU2025a}.
However, they still introduce model-switching overhead and additional complexity in scheduling and memory management.

System-level optimizations can mitigate HBM capacity limits, while the underlying capacity constraint remains and adds system cost. A complementary direction is to extend the memory hierarchy with a storage medium that combines much higher capacity density with sufficiently high read bandwidth.

\subsection{High-Bandwidth Flash as a Capacity Extension}
\label{sec:motivation-hbf}

HBF has been proposed as a high-capacity memory technology for AI inference beyond HBM~\cite{sandiskHBFFactSheet2025}.
As illustrated in Figure~\ref{fig:hbf_concept}, HBF stacks multiple HBF dies and connects them through vertical buses, following a stacked-memory organization similar to HBM.
Each HBF die is based on 3D NAND flash, where multi-layer cell arrays are integrated with logic through vertical bonding.
This flash-based organization gives HBF higher bit density than DRAM-based HBM.

HBF targets high read bandwidth by adapting flash organization for bandwidth-oriented access.
It uses single-level cell (SLC) operation to reduce sensing latency and increases internal parallelism through more sub-arrays, analogous to plane parallelism in 3D NAND.
The stacked organization also provides abundant vertical wiring resources for external bandwidth.
Together, these properties allow an HBF stack to target HBM-comparable read bandwidth with much higher capacity.

However, using HBF in LLM serving introduces challenges because its NAND-flash-based characteristics differ from those of DRAM-based HBM.
Flash reads have higher latency and coarser granularity than DRAM reads, while flash writes are slower and constrained by write endurance.
These characteristics make HBF effective bandwidth depend on access organization.
Long read latency requires enough concurrency or prefetching to avoid bubbles, while page-level access granularity requires data to be consumed in sufficiently large contiguous chunks.
Otherwise, stalls and unused read data reduce effective bandwidth.

To summarize, translating HBF's capacity advantage into performance gains in LLM serving raises challenges at two levels.
At the architecture level, HBM and HBF must be organized to add capacity without reducing effective bandwidth.
At the system level, serving systems must exploit the added capacity to improve serving performance.
The next section addresses these challenges by studying how to architecturally expose large HBF capacity with sufficient bandwidth and how to translate this capacity into system-level benefits for LLM serving.

\section{Design}
\label{sec:design}

\subsection{Architectural Trade-offs of HBF Integration}
\label{subsection:hbf_integration}

\begin{figure}
  \includegraphics[width=\columnwidth]{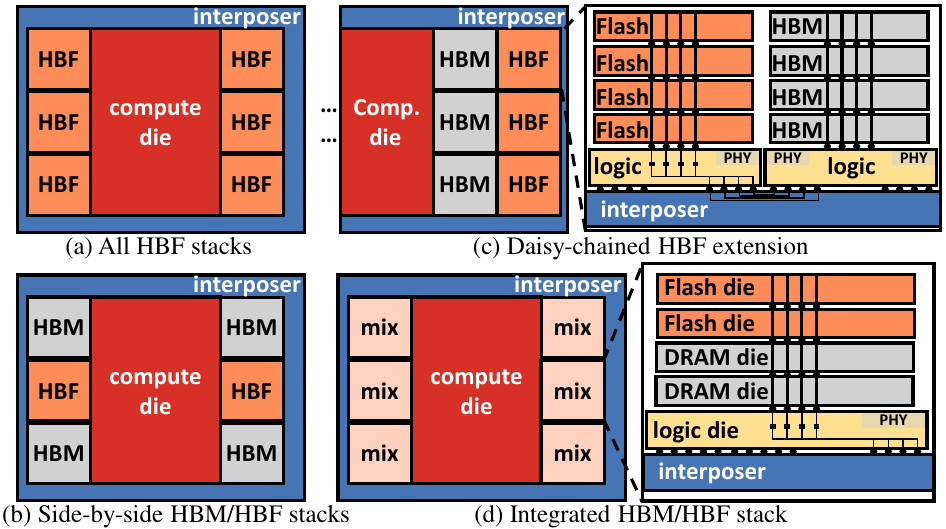}
  \vspace{-1.15em}
  \caption{Representative HBF integration options for GPU memory subsystems.}
  \label{fig:architecture-designs}
  \vspace{-0.5em}
\end{figure}

To satisfy the performance requirements of applications, introducing HBF into the GPU memory subsystem must first ensure that the compute die can obtain \textbf{sufficiently high available bandwidth}. We characterize the bandwidth that the memory subsystem can architecturally expose to the compute die as follows.

\begin{equation}
Bw_{\mathrm{avail}} = 
Bw_{\mathrm{phy}} \cdot U_{\mathrm{intra}} \cdot U_{\mathrm{inter}}.
\label{eq:bw-avail}
\end{equation}

Here, \(Bw_{\mathrm{phy}}\) denotes the physical peak bandwidth provided by the memory media and packaging I/O.
\(U_{\mathrm{intra}}\) captures the utilization within each channel and becomes high when requests match the channel's preferred access granularity, expose sufficient internal parallelism, keep enough requests in flight to hide latency, and avoid excessive control overhead.
\(U_{\mathrm{inter}}\) captures the utilization across channels and becomes high when requests can be evenly distributed across channels with similar access characteristics.

Taking HBM as an example, its high available bandwidth comes from all three factors in Eq.~\ref{eq:bw-avail}.
First, wide I/O, TSV-based stacking, 2.5D packaging, and micro-bump interconnects provide high \(Bw_{\mathrm{phy}}\).
Second, each HBM channel/stack provides rich internal parallelism, a relatively fine access granularity, and modest control overhead. Together with the GPU memory controller's ability to schedule concurrent memory requests, these properties help maintain high \(U_{\mathrm{intra}}\).
Third, HBM channels/stacks are largely homogeneous, making them suitable for fine-grained address mapping and channel interleaving. This allows multiple HBM channels/stacks to be aggregated as a unified bandwidth pool, resulting in high \(U_{\mathrm{inter}}\).

When introducing HBF into the GPU memory subsystem, the three factors in Eq.~\ref{eq:bw-avail} must be considered.
Existing HBF organizations can be classified into the four designs shown in Figure~\ref{fig:architecture-designs}.

\textbf{(a) All-HBF stacks.} This design replaces all HBM stacks with HBF stacks, making HBF directly serve as the main memory of the GPU. The key challenge is that, with no HBM remaining, HBF must serve as the GPU main memory. Without substantial changes to the compute die, the existing \textbf{memory access semantics} are poorly matched to the preferred access granularity of HBF. This makes it difficult for a single HBF channel/stack to sustain high \(U_{\mathrm{intra}}\), and thus prevents \(Bw_{\mathrm{phy}}\) from being fully converted into \(Bw_{\mathrm{avail}}\).

\textbf{(b) Side-by-side HBM/HBF stacks.} This design replaces only a subset of HBM stacks with HBF stacks,
so HBM and HBF are both directly attached to the compute die as \textbf{heterogeneous resources within the same memory tier}. Compared with design (a), the remaining HBM stacks can still serve fine-grained and latency-sensitive memory requests, while HBF can be managed through coarse-grained data movement that better matches its preferred access granularity. This relieves the \(U_{\mathrm{intra}}\) challenge of an all-HBF design. However, because HBM and HBF have different access characteristics, they are difficult to saturate simultaneously under the same workload, shifting the main bottleneck to \(U_{\mathrm{inter}}\).

\textbf{(c) Daisy-chained HBF extension.} This design attaches additional HBF stacks outside the HBM stacks through
a daisy-chain organization, with a multiplexer on the HBM logic base die to reuse or select the I/O path.
Its main advantage is that the direct HBM-to-compute-die path is preserved, so the HBM-side \(U_{\mathrm{intra}}\) and
\(U_{\mathrm{inter}}\) are less likely to be disrupted.
However, HBF accesses must share I/O resources with the HBM path, requiring careful arbitration and scheduling to avoid interfering with critical HBM traffic. In addition, the extra HBF stacks \textbf{increase the package footprint} and may require a larger interposer, raising packaging cost.

\textbf{(d) Integrated HBM/HBF stack.} This design integrates HBM and HBF within each stack, allowing them to
share the TSV-based vertical interconnection, micro-bump I/O, and stack-level interface to the compute die. Compared with design (b), the compute die still sees a set of homogeneous stack-level interfaces, which helps preserve address mapping and interleaving across stacks and maintain high \(U_{\mathrm{inter}}\). Compared with design (a), HBM remains inside each stack, so HBF does not need to serve fine-grained and latency-sensitive
memory requests. The key challenge is the stack-internal I/O sharing, arbitration, and
scheduling. If HBF accesses do not block the critical HBM path, each stack can maintain high \(U_{\mathrm{intra}}\). 

HBF integration exposes a trade-off between \textbf{memory-side design complexity and the available bandwidth delivered to the compute die}. Simpler organizations reduce memory-side complexity by introducing HBF as a coarse-grained stack-level resource, at the cost of possible \(U_{\mathrm{intra}}\) or \(U_{\mathrm{inter}}\) degradation that makes HBM-like \(Bw_{\mathrm{avail}}\) harder to obtain. More integrated organizations move more complexity into the memory side, such as I/O sharing, arbitration, and stack-internal scheduling, and provide more opportunities to preserve the bandwidth utilization conditions of an HBM-based system. For bandwidth-sensitive LLM inference, the integrated HBM/HBF stack offers the strongest option among the four designs because it preserves HBM-like compute-side available bandwidth while integrating HBF capacity without substantially increasing the interposer footprint.

\subsection{HBF-enabled System Design}
\label{subsec:hbf-system-design}

\begin{figure}
  \includegraphics[width=\columnwidth]{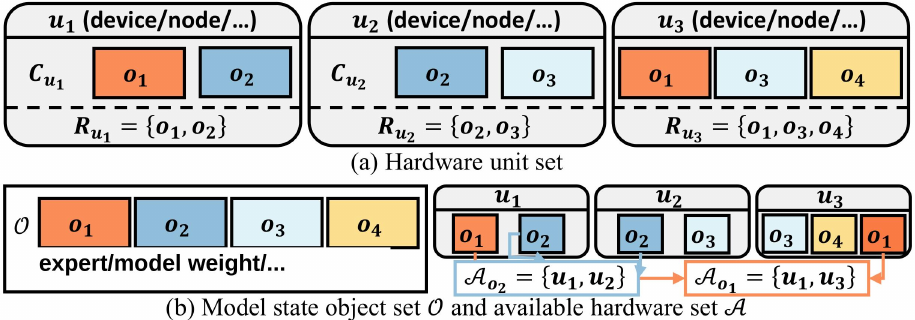}
  \vspace{-1.15em}
  \caption{Capacity-residency model used to reason about locality and load-balancing benefits.}
  \label{fig:scheduling-model}
  \vspace{-0.5em}
\end{figure}

Once HBF is properly organized so that its read bandwidth can be effectively utilized by the compute die, its high capacity density can be translated into system-level performance benefits.
As illustrated in Figure~\ref{fig:scheduling-model}, we model the
capacity benefit as a data-residency-constrained scheduling problem.
Consider a distributed inference system with a set of hardware units
\(\mathcal{U}\), such as devices or nodes, and a set of read-mostly
model-state objects \(\mathcal{O}\), such as MoE expert weights or
model weights.
Each hardware unit \(u \in \mathcal{U}\) has a capacity \(C_u\), and stores a
resident object set \(R_u \subseteq \mathcal{O}\), subject to the capacity
constraint in Eq.~\ref{eq:capacity-constraint}.

\begin{equation}
    \sum_{o \in R_u} \mathrm{size}(o) \le C_u .
    \label{eq:capacity-constraint}
\end{equation}

For any object \(o \in \mathcal{O}\), we define its available hardware set in
Eq.~\ref{eq:available-hardware-set}.

\begin{equation}
    \mathcal{A}_o =
    \left\{
    u \in \mathcal{U}
    \mid
    o \in R_u
    \right\}.
    \label{eq:available-hardware-set}
\end{equation}

Here, \(\mathcal{A}_o\) represents the residency coverage of object \(o\) in the system, i.e., the set of hardware units that can provide this object.
A larger capacity \(C_u\) allows each hardware unit to store a larger resident object set \(R_u\).
As a result, model-state objects can be placed on more hardware units, enlarging their corresponding \(\mathcal{A}_o\).

For a schedulable task \(q\), let its required object set be
\begin{equation}
    \mathcal{D}_q =
    \{o_1, o_2, \ldots\}
    \subseteq \mathcal{O}.
\end{equation}

Executing task \(q\) requires choosing an access or execution location
for each required object \(o_i \in \mathcal{D}_q\), and the candidate
locations for each object are determined by its corresponding
\(\mathcal{A}_{o_i}\).
Therefore, the scheduling space of a task is jointly determined by the
available hardware sets of all its required objects.

Under this model, the capacity benefit of HBF can be summarized in two
ways.
\textbf{First, a larger \(\mathcal{A}_o\) improves execution locality.}
For a task that depends on multiple objects, if the available hardware
sets of these objects have larger overlap on the same devices or nodes, the
required objects are more likely to be accessed or
executed within the local communication domain.
This reduces cross-node communication.
\textbf{Second, a larger \(\mathcal{A}_o\) improves load balancing.}
For tasks that depend on object \(o\), placing \(o\) on more hardware
units gives the scheduler a larger candidate set from which to choose a
less loaded device or communication domain, reducing queuing delay and
tail latency caused by hot objects.

For one specific model-state object \(o\), if the capacity \(C_u\) of each hardware
unit increases linearly, the system can place more replicas of this
object.
Thus, the average replica count \(r_o\) also increases approximately
linearly with capacity.
A larger replica count increases the probability \(\rho_o\) that object
\(o\) can be found within the local communication domain.
Equivalently, the probability \(P_{\mathrm{remote}}(o)\) that accessing
\(o\) still requires cross-node access decreases approximately linearly
before local coverage saturates.
Therefore, the number of remote accesses, cross-node traffic, and the
communication pressure caused by remote accesses can be approximated as
decreasing \textit{linearly} with capacity.
On the other hand, if tasks depending on the same object can be evenly
distributed across its replicas, the average load per replica is
inversely proportional to the replica count.
Since the replica count grows approximately linearly with capacity, the
per-replica load pressure of a hot object decreases approximately
\textit{inversely} with capacity.
Increasing capacity expands object coverage, which approximately
reduces cross-node communication pressure and reduces per-replica load
pressure in proportion to \(1/C_u\), as summarized in Eq.~\ref{eq:capacity-benefit}.
\begin{equation}
\begin{aligned}
    &P_{\mathrm{remote}}(o)
    \approx 1-\rho_o,
    &\rho_o &\propto r_o \propto C_u, \\
    &\mathrm{Load}_{\mathrm{per\ replica}}(o)
    \propto \frac{1}{r_o}
    \propto \frac{1}{C_u}.
\end{aligned}
\label{eq:capacity-benefit}
\end{equation}
This model provides a trend between capacity,
communication pressure, and load pressure.

\begin{figure}
  \includegraphics[width=\columnwidth]{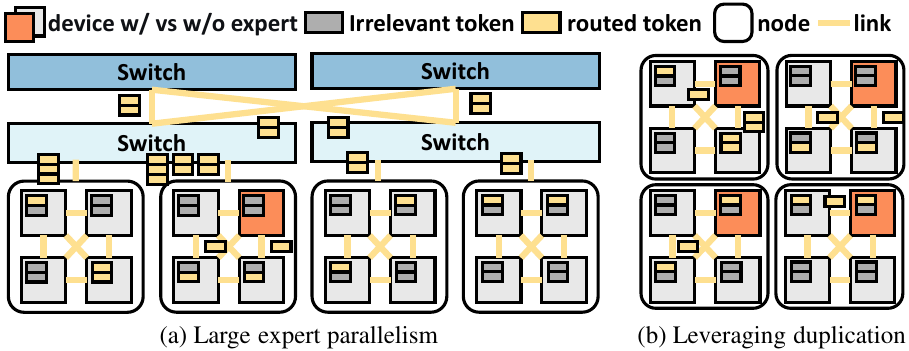}
  \vspace{-1.15em}
  \caption{How additional capacity enables expert replication and reduces remote expert communication in MoE serving.}
  \label{fig:moe-benefits}
  \vspace{-0.5em}
\end{figure}

\textbf{In MoE inference}, expert weights are the model-state objects.
\begin{equation}
    \mathcal{O} =
    \{E_1, E_2, \ldots, E_K\}.
\end{equation}
For a token or token group \(q\), the router selects a set of experts.
\begin{equation}
    \mathcal{D}_q =
    \{E_{i_1}, E_{i_2}, \ldots, E_{i_k}\}.
\end{equation}
For each expert \(E_i\), \(\mathcal{A}_{E_i}\) represents the devices or
nodes on which this expert resides.
Figure~\ref{fig:moe-benefits} illustrates how additional capacity
benefits MoE serving. Large-scale expert parallelism distributes
experts across devices and nodes, which introduces remote
dispatch/combine traffic when a token is routed to non-local experts.
With additional HBF capacity, more experts can be replicated within
the local communication domain, enlarging the corresponding
\(\mathcal{A}_{E_i}\).
This makes the experts required by a token more likely to be accessed or
executed within the local communication domain, reducing remote expert
accesses and cross-node traffic.
Moreover, when hot experts have more replicas, the scheduler can choose a
less loaded replica, mitigating compute and communication imbalance
caused by expert hotspots.
Therefore, in the MoE case, the main capacity benefit is to enlarge the
available hardware set of experts, reduce remote expert communication,
and improve the load distribution of hot experts.

\begin{figure}
  \includegraphics[width=\columnwidth]{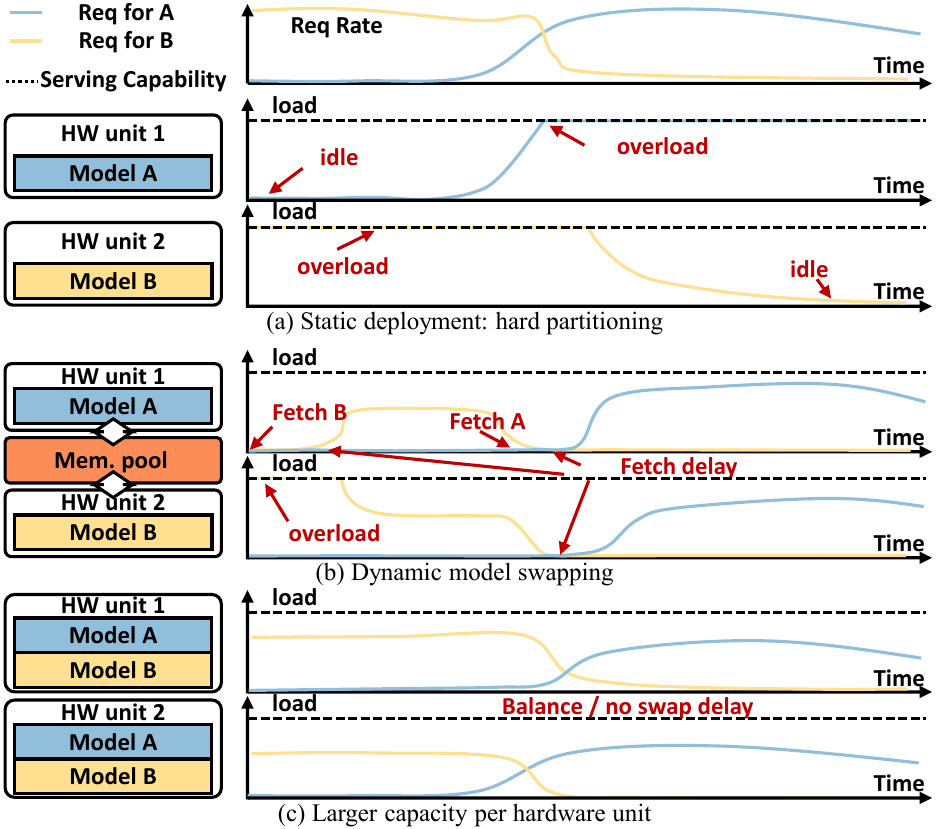}
  \vspace{-1.15em}
  \caption{How larger per-device capacity reduces model loading and enables better load balance in multi-model serving.}
  \label{fig:multi-model-serving}
  \vspace{-0.5em}
\end{figure}

\textbf{In multi-model serving}, the weights of different models are the
model-state objects.
\begin{equation}
    \mathcal{O} =
    \{M_1, M_2, \ldots, M_L\}.
\end{equation}
For a request or request batch \(q\) targeting model \(M_j\), the required
object set is usually
\begin{equation}
    \mathcal{D}_q = \{M_j\}.
\end{equation}
Figure~\ref{fig:multi-model-serving} illustrates the multi-model
serving case. When capacity is limited, models must be hard-partitioned
or dynamically loaded and evicted. With larger per-device capacity,
more model weights can remain resident across hardware units,
expanding \(\mathcal{A}_{M_j}\) and giving the scheduler more placement
choices.
This allows the scheduler to assign requests from a larger candidate set,
avoid overloaded devices, reduce queuing delay and tail latency, and
reduce the need for dynamic model loading and eviction.
Communication overhead may also exist in this case, while request dispatch
traffic is usually small and occurs once per request, so it is less likely
to dominate performance.
Therefore, in multi-model serving, the main capacity benefit is to expand
model replica coverage, distribute requests for the same model across
more hardware units, reduce per-replica load pressure, and improve
scheduling flexibility.

HBF-enabled system design can be viewed as expanding the
available hardware set \(\mathcal{A}_o\) of model-state objects by
increasing capacity, thereby relaxing data-residency constraints in
workload scheduling.
For MoE inference, this mainly improves expert access locality and
reduces cross-node communication.
For multi-model serving, this mainly expands model replica coverage and
improves request load balancing.

\section{Evaluation}
\label{sec:evaluation}

\subsection{Methodology}
\label{sec:evaluation_methodology}

\paragraph{\textbf{Evaluation scope.}}
We evaluate the system-level benefits of additional memory capacity in
two representative LLM serving scenarios, MoE serving and multi-model
serving. We study each scenario separately to isolate its capacity
benefit.

\paragraph{\textbf{HBF capacity assumption.}}
Our evaluation isolates the system-level benefit of HBF-backed capacity
under the bandwidth-preserving integration assumption discussed in
Section~\ref{subsection:hbf_integration}. We model HBF as additional
device-local capacity for read-mostly model-state objects, while assuming that
the integrated HBM/HBF organization can preserve the compute-side
available bandwidth through independent stack-level bandwidth, page-granular
weight layout, prefetch commands, and latency-hiding buffers. Therefore,
the experiments should be interpreted as the capacity benefit achievable
when HBF accesses do not reduce the effective bandwidth of the
HBM-resident execution path. If HBF is exposed through a slower
off-package path, if weight reads cannot be organized at page granularity,
or if HBF traffic interferes with the critical HBM path, the realized
benefit would be smaller. We do not place the mutable KV cache or runtime
buffers in HBF in this study. They remain HBM-resident, while HBF
is used for read-mostly model weights and expert weights.

\paragraph{\textbf{Simulation framework.}}
We implement an LLM serving simulator for both experiments. The simulator supports
\textbf{(i) Workload traces.}
synthetic and trace-driven workloads with request arrival times,
input/output lengths, target models, and expert selections;
\textbf{(ii) Operator performance.}
operator compute and memory-access volumes derived from model
configurations, with latency estimated by a roofline model
parameterized by compute throughput and effective memory bandwidth;
\textbf{(iii) Serving dynamics.}
event-driven request batching, expert/model placement, replica
scheduling, and model loading/eviction;
\textbf{(iv) Communication and data movement.}
scale-up, scale-out, and memory-tier data movement with link resources
parameterized by bandwidth and latency; and
\textbf{(v) Capacity and residency.}
device memory capacity and model-state residency, with HBF abstracted
as additional device-local capacity under the compute-side
available-bandwidth assumption in
Section~\ref{subsection:hbf_integration}.

\subsection{MoE Serving Evaluation}
\label{sec:moe_evaluation}

\paragraph{\textbf{Setup.}}
Table~\ref{tab:moe_setup} summarizes the MoE serving setup. We evaluate
Qwen3-235B-A22B under a PD-separated serving configuration and vary the
resident expert capacity to study how HBF-backed capacity affects expert
replication, locality, and latency.

\begin{table}[t]
\centering
\caption{MoE serving evaluation setup.}
\label{tab:moe_setup}
\vspace{-0.6em}
\small
\renewcommand{\arraystretch}{1.0}
\setlength{\tabcolsep}{3.5pt}
\begin{tabular}{|p{0.28\columnwidth}|p{0.64\columnwidth}|}
\hline
\textbf{Category} & \textbf{Configuration} \tabularnewline
\hline
Model & Qwen3-235B-A22B, 94 layers, 128 routed experts, top-8 experts per token. \tabularnewline
\hline
Cluster & 4 H100-class nodes, including 2 prefill nodes and 2 decode nodes, 8 GPUs per node. \tabularnewline
\hline
Memory & 80~GB HBM + 512~GB HBF per GPU for resident expert replicas. \tabularnewline
\hline
Interconnect & 900~GB/s scale-up within node; 400~GB/s scale-out across nodes. \tabularnewline
\hline
Parallelism & DP for attention/projection, EP for MoE FFN, and no tensor parallelism. \tabularnewline
\hline
Placement/\newline scheduling & EPLB-style hierarchical expert replication~\cite{deepseek_eplb}, with expert-load-aware runtime scheduling~\cite{yu2025metro}. \tabularnewline
\hline
Workload & Poisson arrivals, with log-normal prompt/output lengths of mean 1024/64 tokens. \tabularnewline
\hline
Expert access & Hotset-Zipf distribution, where 16 hot experts cover 90\% of expert demand. \tabularnewline
\hline
\end{tabular}
\vspace{-0.8em}
\end{table}

\begin{figure*}[t]
\centering
\includegraphics[width=\textwidth]{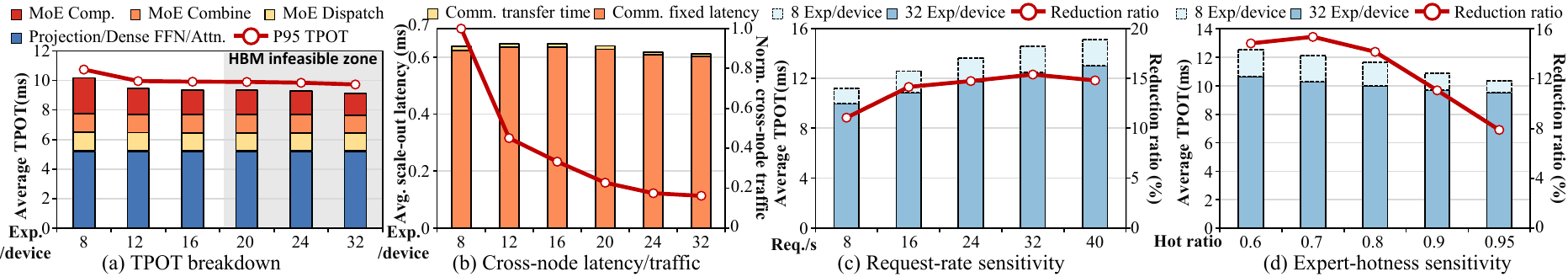}
\vspace{-1.25em}
\caption{Impact of HBF-backed expert capacity on MoE decoding latency, communication locality, and workload sensitivity.}
\label{fig:exp_moe_benefits}
\vspace{-0.7em}
\end{figure*}

\paragraph{\textbf{Takeaway 1. HBF-backed capacity improves MoE decoding latency.}}
Figure~\ref{fig:exp_moe_benefits}(a) first evaluates the capacity
sweep under a fixed request rate of 4~req/s. In this model, one resident
expert replica consumes about 3.55~GB per device when counted across
the 94 MoE layers. Therefore, 8, 12, 16, 20, 24, and 32 resident experts
per device consume about 28.4, 42.6, 56.8, 71.0, 85.2, and 113.5~GB,
which correspond to 35.5\%, 53.2\%, 71.0\%, 88.7\%, 106.5\%, and
141.9\% of an 80~GB HBM device. Without capacity expansion, even
moderate expert duplication would leave very limited HBM space for the
KV cache and runtime reserve buffers, and the 24- and 32-expert points
would exceed the HBM capacity. HBF-backed capacity makes these duplicated
expert placements feasible without squeezing HBM-resident execution
state. With this additional residency space, hot experts can be duplicated
across more devices, which reduces the bottleneck device's active-expert
count and shortens the MoE compute barrier. As a result, increasing the
resident expert capacity from 8 to 32 experts per device reduces average
TPOT from 10.20~ms to 9.13~ms, a 10.5\% reduction, while p95 TPOT
decreases from 10.74~ms to 9.73~ms, a 9.4\% reduction.

\paragraph{\textbf{Takeaway 2. Duplicated experts reduce cross-node traffic.}}
Figure~\ref{fig:exp_moe_benefits}(b) reports the normalized cross-node
expert traffic under the same 4~req/s setting. As the resident
expert capacity increases from 12, 16, 20, 24, to 32 experts per
device, the normalized cross-node traffic decreases to 45.1\%,
33.3\%, 22.8\%, 17.4\%, and 16.1\% of the 8-expert baseline,
respectively. This is because a larger resident expert budget gives the
placement policy and runtime scheduler more local replica candidates, so
more dispatch/combine transfers stay within the local node. This trend is
consistent with the capacity-locality model in
Section~\ref{subsec:hbf-system-design}. Since scale-out communication is a
barrier determined by the slowest transfer path, lower aggregate traffic may
leave latency nearly unchanged when the slowest path remains similar.

\paragraph{\textbf{Takeaway 3. Capacity benefit depends on request rate.}}
Figure~\ref{fig:exp_moe_benefits}(c) compares 8 and 32 experts per
device under different request rates. The larger-capacity setting
consistently lowers average TPOT, and the magnitude of improvement is
decode-batch-regime dependent. As the request rate increases from 8 to
32~req/s, the reduction grows from 11.1\% to 15.4\%, because larger
decode batches expose more simultaneous expert activations and make the
bottleneck device's active-expert count more important. At 40~req/s, 
the reduction slightly decreases to 14.8\%. This suggests
that the workload is moving into a denser decode-batch regime, where
more simultaneous tokens partially amortize fixed communication and
scheduling costs. Therefore, the capacity benefit is strongest when
the decode batch is large enough to expose bottleneck-device imbalance,
while still sparse enough for this imbalance to remain visible after
batch-level averaging.

\paragraph{\textbf{Takeaway 4. Expert hotness changes capacity benefit.}}
Figure~\ref{fig:exp_moe_benefits}(d) varies the hot-expert ratio while
keeping the number of hot experts fixed. The 32-expert setting lowers
average TPOT across all hotness levels. The reduction is largest
for diffuse-to-moderately hot workloads, reaching 14.8\% at hot ratio
0.6 and 15.4\% at 0.7. As the access pattern becomes more concentrated,
the reduction decreases to 11.1\% at hot ratio 0.9 and 7.9\% at 0.95.
This behavior follows from the capacity-locality and load-balancing
model. When expert accesses are more diffuse, each decode step can
activate a wider and less predictable expert set, so the 8-expert
placement may still concentrate active experts on a subset of devices.
Extra resident capacity provides broader replica coverage and gives
the scheduler more choices to avoid bottleneck active-expert counts.
When the access pattern is highly concentrated, each decode step activates
a narrower and more predictable expert set. The baseline placement therefore
has less remaining active-expert imbalance for additional replicas to remove.

\subsection{Multi-Model LLM Serving} 

\paragraph{\textbf{Experimental setup.}}

We evaluate GPU memory capacity in a 4-GPU PD-separated setup with two
prefill GPUs and two decode GPUs, organized as two placement groups
with one GPU on each side. Because model weights must be resident on
both sides before serving, we report model-weight residency at the
placement-group level. The baseline follows NVIDIA H100 parameters,
80~GB HBM per GPU, H100 compute throughput, and PCIe Gen5 x16 H2D
offload bandwidth modeled as 64~GB/s unidirectional bandwidth.

HBF-backed capacity provides up to 512~GB of on-device model-weight
residency per placement group, or 6.4$\times$ the 80~GB HBM baseline.
We sweep the per-placement-group capacity from 1$\times$ to 6.4$\times$
while fixing KV-cache space, compute throughput, H2D bandwidth, and KV
transfer bandwidth. Scheduling uses Prism with KVPR-based dynamic model
placement. For model replication (Takeaway 2), hot models are selected
by request popularity and proactively replicated across prefill GPUs;
other models follow the Prism placement plan.

For Takeaways 1 and 2, we use a production trace with 1,500 requests
over 600 seconds, 27 heterogeneous models (1B--40B parameters),
Zipf-distributed popularity ($\alpha \approx$ 1.49), 2.5 req/s average
arrival rate, and about 620~GB aggregate FP16 weight footprint. For
Takeaway 3, we use synthetic traces with 18 models (total
$\approx$ 507~GB) to control arrival rate.

\paragraph{\textbf{Takeaway 1. Capacity scaling eliminates model loading overhead.}}

When device memory cannot accommodate all model weights simultaneously, serving systems must dynamically load and evict models on demand, introducing substantial wait latency that dominates Time-To-First-Token (TTFT). In the production trace, increasing GPU memory capacity progressively eliminates this overhead. Mean TTFT decreases from 196.1 ms at 1× capacity to 6.8 ms at 4× capacity, a 28.7× reduction. Concurrently, model activation events fall from 212 to zero. Beyond the 4× threshold where all 27 models fit simultaneously in memory, further capacity increases yield no additional TTFT improvement from the loading-elimination mechanism alone, indicating an all-fit boundary.

The improvement originates entirely from eliminating model loading wait time on the request's serving path. In the PD-separated architecture, both prefill and decode nodes require model weights to be resident before processing. At 1× capacity (80~GB per placement group), only a fraction of the 27 models can remain loaded; requests targeting evicted models must wait for Host-to-Device (H2D) weight transfer, which dominates end-to-end latency. As capacity grows from 1× to 3×, the working set of resident models expands, reducing model activation frequency from 212 to just 4 events. At 4× capacity, the total weight footprint (620~GB) fits within the effective model-placement capacity across two placement groups (\(2 \times 4 \times 80\)~GB = 640~GB), entirely removing load-on-demand overhead.

\begin{figure}[H]
\centering
\includegraphics[width=\linewidth]{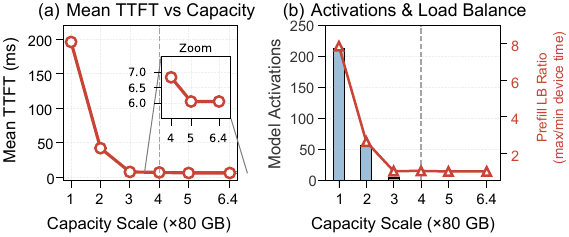}
\vspace{-1.15em}
\caption{Capacity scaling in multi-model serving. All-fit capacity removes
loading, and surplus capacity balances prefill load.}
\vspace{-1.15em}
\label{fig:takeaway12}
\end{figure}

\paragraph{\textbf{Takeaway 2. Surplus capacity enables weight replication for load balancing.}}

Beyond the all-fit threshold, additional memory capacity permits replication of popular model weights across multiple prefill GPUs to balance computational load. In the baseline (no replication) configuration, each model resides on exactly one prefill GPU; requests targeting popular models queue behind each other on a single device, creating load imbalance. With static replication, the hottest models are duplicated across both prefill GPUs, allowing incoming requests to be routed to the less-loaded device.

As shown in Figure~\ref{fig:takeaway12}(b), the prefill load balance ratio (defined as the ratio of the most-utilized to least-utilized prefill device running time) drops from 7.87$\times$ at 1$\times$ capacity to approximately 1.03 at 4$\times$ capacity as model loading overhead disappears; with replication enabled at capacity levels of 5$\times$--6.4$\times$, the ratio reaches 1.0. This improved load balance translates directly into reduced prefill queue wait time. With replication, the mean TTFT decreases further from 6.83~ms (at 4$\times$ without replication) to 6.05~ms, a modest but consistent 11\% improvement enabled by the surplus memory capacity that HBF provides beyond the all-fit threshold.

Memory capacity beyond the all-fit point can therefore be used for replication-based load balancing, providing an additional performance optimization that complements the primary benefit of model residency.

\paragraph{\textbf{Takeaway 3. Greater active model set volatility amplifies capacity benefit.}}

\begin{figure}[t]
\centering
\includegraphics[width=\linewidth]{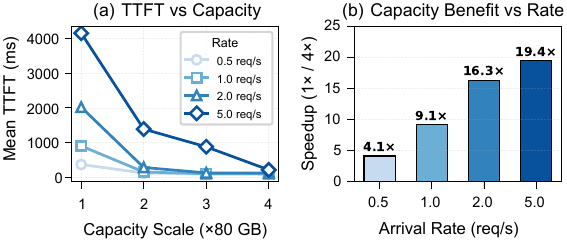}
\vspace{-0.9em}
\caption{Request-rate sensitivity in multi-model serving. Capacity speedup
grows with arrival rate.}
\vspace{-0.7em}
\label{fig:takeaway3}
\end{figure}

Figure~\ref{fig:takeaway3} shows that expanded memory capacity becomes more valuable as the active model set changes more rapidly. A higher request arrival rate increases the frequency at which different models are targeted per unit time, causing more rapid fluctuations in the set of models that must be simultaneously resident. We vary the request arrival rate from 0.5 to 5.0 req/s while holding all other parameters fixed, and measure the TTFT speedup achieved by scaling from 1× to 4× capacity. The speedup increases with request rate. At 0.5 req/s, capacity scaling yields a 4.1× TTFT improvement (368 ms → 90 ms); at 1.0 req/s, 9.1× (902 ms → 99 ms); at 2.0 req/s, 16.3× (2,020 ms → 124 ms); and at 5.0 req/s, 19.4× (4,151 ms → 214 ms). At 4× capacity, where all models fit, TTFT remains consistently low (90–214 ms) regardless of rate. This indicates model loading overhead as the source of the amplified penalty at low capacity.

The amplification arises from the interaction between active model set volatility and loading latency. At low rates (0.5 req/s), requests are temporally spread and the active model set changes slowly. Each incoming request typically waits only for its own model to load, with 238 model activation events distributed over shallow queue depths, limiting TTFT to 368 ms. At higher rates (5.0 req/s), the active model set fluctuates rapidly. Requests targeting diverse models arrive every 200 ms on average, and when a model requires loading (taking 2--5 seconds for large models), subsequent requests queue behind the loading operation. Although the total number of model activation events is actually lower at high rates (103 vs 238, since models remain loaded between closely-spaced requests), each activation event blocks proportionally more concurrent requests, yielding higher mean TTFT. This finding has direct implications for HBF deployment. In production multi-model environments where the active model set fluctuates rapidly, aggregate request rates in the 1--5 req/s range can make the capacity bottleneck amplify TTFT by 9--19×, while 512~GB of HBF-backed on-device capacity removes the volatility-dependent loading penalty in this setup.

\section{Conclusion}
\label{sec:conclusion}

This report studies how High-Bandwidth Flash (HBF) could be used as a
capacity-oriented extension to the memory hierarchy of LLM serving systems.
The motivation is that serving workloads are growing along multiple data
dimensions. Larger models increase weight footprint, longer contexts increase
KV-cache footprint, and broader model families increase the amount of model
state that a service must keep available. These trends make HBM capacity a
primary constraint, especially for large MoE models and multi-model
serving.

At the architecture level, adding HBF capacity must preserve the effective
bandwidth available to the compute die. We therefore
compare several HBM/HBF integration options and identify an integrated HBM/HBF
stack as a promising organization. HBM continues to serve fine-grained,
latency-sensitive accesses, and HBF provides additional read-mostly capacity
for coarse-grained model-state objects. This design direction preserves the
compute-side bandwidth abstraction of an HBM-based system while creating a
larger residency space for weights and experts.

At the system level, HBF-backed capacity is useful when serving performance is
limited by model-state residency. In MoE serving, additional capacity can hold
more expert replicas, improving expert locality and reducing cross-node expert
communication. In our Qwen3-235B-A22B study, increasing resident expert capacity
reduces average TPOT and lowers cross-node expert traffic by giving the
scheduler more local expert choices. In multi-model serving, additional
capacity reduces model loading and eviction. Once the active model set fits in
device-local capacity, TTFT decreases substantially; surplus capacity can then be used to
replicate hot models and improve load balance.

HBF complements HBM as a capacity-oriented extension. Its value comes from combining
hardware support that exposes high-capacity read bandwidth with serving policies
that convert the extra residency space into locality, load balancing, and
reduced data movement. In LLM serving, HBM should continue to hold
bandwidth-critical execution state, while HBF expands the resident set of
read-mostly model-state objects.

\bibliographystyle{unsrtnat}
\bibliography{references}

\end{document}